# A systematic investigation of electric field nonlinearity and field reversal in low pressure capacitive discharges driven by sawtooth-like waveforms

Sarveshwar Sharma[1,2], Nishant Sirse[3], Miles M Turner[4] and Animesh Kuley[5]

[1]Institute for Plasma Research, Gandhinagar-382428, India
[2]Homi Bhabha National Institute, Anushaktinagar, Mumbai-400094, India
[3]Institute of Science and Research and Centre for Scientific and Applied Research, IPS Academy, Indore-452012, India
[4]School of Physical Sciences and National Center for Plasma Science and Technology, Dublin City University, Dublin 9, Ireland
[5]Department of Physics, Indian Institute of Sciences, Bangalore-560012, India

Email: nishantsirse@ipsacademy.org

**Abstract**

Understanding electron and ion heating phenomenon in capacitively coupled radio-frequency plasma discharges is vital for many plasma processing applications. In this article, using particle-in-cell simulation technique we investigate the collisionless argon discharge excited by temporally asymmetric sawtooth-like waveform. In particular, a systematic study of the electric field nonlinearity and field reversal phenomenon by varying the number of harmonics and its effect on electron and ion heating is performed. The simulation results predict higher harmonics generation and multiple field reversal regions formation with an increasing number of harmonics along with the local charge separation and significant displacement current outside sheath region. The field reversal strength is greater during the expanding phase of the sheath edge in comparison to its collapsing phase causing significant ion cooling. The observed behavior is associated with the electron fluid compression/rarefaction and electron inertia during expanding and collapsing phase respectively.

# 1. Introduction

Capacitively coupled plasma (*CCP*) discharges operated in the low-pressure collisionless regime are the mainstream tool for microelectronics device fabrication. These devices are utilized in critical manufacturing steps such as thin film deposition by the process called Plasma Enhanced Chemical Vapour Deposition (*PECVD*) or removal of one or more materials selectively as performed by the reactive ion etcher (*RIE*) [1]. Both processes involve precise optimization and control of plasma properties and surface conditions. For instance, independent control of the flux and energy of the charge particles impinging on the wafer is crucial for improving plasma processing rates with lesser damage to the substrate [2]. In addition, plasma uniformity is an important consideration for large-area substrate processing.

In *CCP* discharges, the plasma properties including electron energy distribution functions (*EEDF*) that drive plasma processing chemistries and surface processes are mostly governed by electron and ion power deposition. For collisionless conditions, electron power absorption from the applied RF field is mostly occurs during the oscillating phase of the sheath edge near to the electrodes, whereas ions are accelerated in the time-averaged sheath potential. In single frequency *CCP* discharges operated by sinusoidal waveform, the electron heating during the expanding phase of the sheath edge has been widely investigated using computer simulation [2-19], experimentally [20-29] and theoretically [30-37]. For specific operating conditions, it has been observed that the electrons are also heated during the collapsing phase of the sheath edge [38-40]. This heating effect is due to the fact that the electron diffusion cannot follow fast sheath collapse and thus local field reversal is formed that further accelerates electrons. By performing the electric field diagnostics using electron beam measurements Sato and Liebermann [38] reported the field reversal phenomenon in argon plasma discharge. Vendor and Boswell [39] performed *PIC* simulation studies and investigated the formation of field reversal in hydrogen *CCP* discharge. Schulze *et. al.* [40] employed phase resolved optical emission spectroscopy to investigate the electric field reversal in both asymmetric single and symmetric dual-frequency *CCP* discharges. The work conclude that the field reversal causes significant and a general source of electron heating during the sheath collapse in rare as well as in molecular gas discharges. The formation of field reversal during the expanding phase of the sheath edge was first reported by Sharma and Turner [6]. In their *PIC* simulation studies it was observed that the field reversal is formed during sheath expansion for higher current density (~100 A/m$^2$) and its strength is greater when compared to collapsing phase field reversal strength. Electron fluid compression and rarefaction were found to be one of the main reasons for this phenomenon. Furthermore, it was noticed that the electron plasma waves originate from the region near to local field reversal. In another *PIC* study by Sharma and Turner [41], multiple field reversals were observed in dual-frequency *CCP* discharge. Field reversals were also observed at high pressure conditions and in electronegative plasmas caused by electron drag force and double layer formation respectively [42-51].

The above studies were focused on the *CCP* discharges excited by the sinusoidal waveforms. In the last decade, *CCP* systems excited by non-sinusoidal waveforms [52-65] have shown a promising way to control the electron and ion heating asymmetry in the discharge that is useful in many plasma processing applications. These waveforms are known as "tailored waveform" produced by superimposing higher harmonics in the sinusoidal waveform. Both amplitude asymmetric [66, 67] and temporal asymmetric [56, 57, 64] waveforms have been utilized. It is observed that the temporally asymmetric waveforms offer advantages in terms of generating heating asymmetry without affecting the ion energy asymmetry. Several studies have been published employing temporally asymmetric sawtooth-like waveform to study the asymmetric heating response and its effect on electron and ion energy distribution functions at low and very high frequency [61-63]. More recently [64], the effect of increasing number of harmonics (*N*) in sawtooth-like waveform on ion energy deposition is systematically presented.

We follow our recent study and extend the previous work to investigate the nonlinearity on the oscillating sheath edge and electron beam with systematic variation in *N*. The field reversal has been observed earlier but never reported in the *CCP* discharges excited by tailored waveform. It is equally important to reveal the underlying electron and ion heating mechanism in such discharges. The present manuscript report particle-in-cell (*PIC*) simulation study of field reversal phenomenon in the *CCP* discharges excited by temporally asymmetric sawtooth-like waveform and its effect on electron and ion heating. The number of harmonics (*N*) contained in the sawtooth-like waveform is systematically varied from perfectly sinusoidal waveform (*N*=1) to *N*=12. The study is performed for a fixed gas pressure of 5 mTorr and by keeping the current density constant at 50 A/m$^2$.

The paper is structured as follows: In section II, the details of simulation scheme along with the initial parameters is described. Sec. III, discuss the simulation results and finally the summary and conclusions of the work are presented in Sec. IV.

## 2. Simulation Scheme and parameters

In the present study, we have utilized a particle-in-cell (*PIC*) simulation scheme in conjunction with the Monte Carlo Collision Scheme [68, 69]. The simulation was carried out in one spatial dimension and three velocity dimensions (*1D-3V*) using a thoroughly tested and validated code [70], which has been widely employed in the literature to investigate various problems in capacitively coupled discharges [9-11,14,16,17,63,64,71-73,76,77]. This code is versatile enough to handle both current and voltage-driven modes, and for all sets of simulations, we opted to use the former. To excite the plasma, we selected a sawtooth-like current waveform, which is mathematically represented by the following equation [56]:

$$J_{rf}(t) = \pm J_0 \sum_{k=1}^{N} \frac{1}{k} \sin(k\omega_{rf} t) \quad \text{------------ (1)}$$

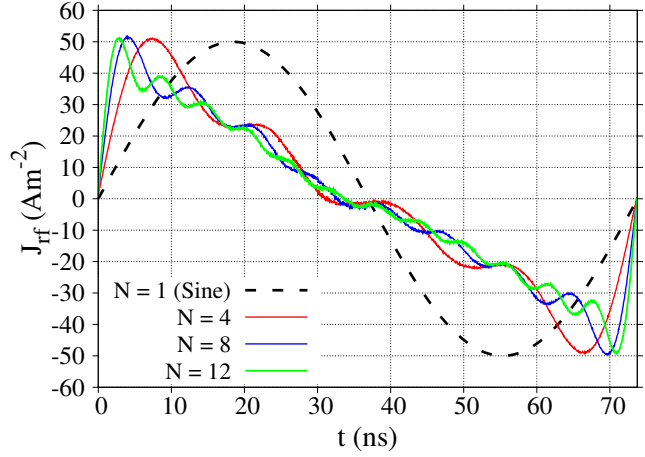

Figure 1. Current density profile for 4 different values of $N$ (1(sinusoidal), 4, 8 and 12) at the powered electrode.

.

Equation (1) describes the sawtooth current waveform, where $J_0$ represents the current density amplitude, and $\omega_{rf}$ is the fundamental angular frequency. The positive and negative signs in Eq. (1) correspond to the "sawtooth-down" and "sawtooth-up" waveforms, respectively. In our current simulation, $N$ denotes the number of harmonics contained in the sawtooth waveform, and we explored its variation from 2 to 12. Figure 1 displays the applied current density profiles at the powered electrode for different $N$ values i.e., 1(sinusoidal), 4, 8, and 12 obtained from the simulation. The total current density amplitude ($J_0$) remained constant at 50 A/m² across all simulations, while the fundamental frequency was consistently set at 13.56 MHz. As shown in In Figure 1, it is evident that the current profile exhibits temporal symmetry when $N = 1$, corresponding to the sinusoidal waveform. However, as the number of harmonics increases, the rise time (from 0 to 50 A/m²) significantly decreases from 12.35 to 2.79 ns for $N = 2$ to $N = 12$, respectively. The current waveforms are applied at the powered electrode ($L$=0 mm), while the grounded electrode is positioned at $L$=60 mm. For the simulations, an external capacitor is not taken into account. The *DC* self-bias generated at the powered electrode is a result of the temporal asymmetry, arising in a self-consistent manner from the simulation based on the charge imbalance at the powered and grounded electrodes.

The simulation is focused on electropositive argon plasma, where the gas is uniformly distributed in the discharge region at a temperature of 300 K. Throughout the simulation, the gas pressure remains constant at 5 mTorr. A comprehensive range of reactions is taken into account, including ion-neutral interactions (elastic, inelastic, and charge exchange) and electron-neutral interactions (elastic, inelastic, and ionization). Additionally, the simulation considers the presence of two metastable states, $Ar^*$ and $Ar^{**}$, which represent excited states of argon at 11.6 eV ($Ar^*$-$3p^54s$) and

13.1 eV ($Ar^{**}$-$3p^54p$) respectively. This simulation incorporates several additional processes that influence the population and de-population of excited states. These processes encompass de-excitation, super elastic collisions, metastable pooling, and multi-step ionizations. To ensure accuracy and reliability, the collision cross sections for all the reactions are sourced from well-established and thoroughly tested references [14,73,78]. The simulation assumes perfectly absorbing electrodes, without taking into account electron or ion-induced secondary electron emissions (*SEE*). This choice is well-suited for the current conditions, given the low operating pressure and correspondingly lower discharge voltage. Consequently, the presence of *SEE* is not expected to significantly affect the dynamics of the plasma.

In order to achieve a steady-state solution, all simulations sets were run for over 5000 RF cycles. The stability and accuracy of the simulation were carefully maintained by selecting suitable grid size ($\Delta x$) and time step ($\Delta t$) to effectively resolve the Debye length ($\lambda_{De}$) and electron plasma frequency ($\omega_{pe}$), respectively [68].

## 3. Results and Discussions

Figure 2 shows the electric field spatiotemporal distribution in the discharge for different values of *N* from 1 (sine) up to 12. All the graphs shown in figure 2 are plotted for one RF period and averaged over 100 RF cycles after achieving the steady state. The graphs are normalized from $-1 \times 10^4$ V/m to $+1 \times 10^4$ V/m i.e., presented on the same color bar scale for the comparison and better visualization of field reversal regimes. The powered electrode is positioned at *L*=0 mm, while the grounded electrode is located at *L*=60 mm. Figure 2 (a) depicts, for sinusoidal case (*N*=1), the electric field is predominantly confined within the sheath region with minimal presence in the bulk plasma. The sheath edges on both the powered and grounded electrodes are smooth and both the sheaths are symmetric in this case. As the value of *N* increases, we observe that the sheath electric field starts penetrating in bulk plasma from near to the initial phase of expanding sheath edge (*t*~5 ns) at grounded electrode. This electric field penetration is due to the formation of energetic electron beams that originated after electrons gain energy with oscillating sheath edge position. It is noticed that as the value of *N* increases the plasma distribution becomes asymmetric with bigger sheaths near to the grounded electrode in comparison to the powered electrode. This asymmetric nature of the applied saw-tooth waveform is responsible for the higher energy gained by the electrons from the sheath at the grounded electrode. In addition to asymmetry, multiple electron beams generation and penetration is observed for higher values of *N* (for example, indicated in (*h*, N=8)). It is also observed that the sheath is perfectly sinusoidal for *N*=1, however, as the value of *N* increases high frequency oscillations are observed on the instantaneous sheath edge position. These high frequency oscillations are analogue to the multiple frequency operated *CCP* discharges where higher frequency is superimposed on the low frequency for an independent control of ion flux and ion energy [2,18,41,79-84]. It is evident that these high frequency oscillations at higher values of *N* are responsible for an enhancement in the conjugate sheath velocity, which in turn transfer higher energy to the electrons which are interacting with the oscillating sheath as described by the hard wall model [30] and therefore the electron beam penetration is observed (effect is weaker at lower *N* and stronger at higher *N*). These electron beams carry energy from the sheath to bulk plasma resulting in an increase in the plasma density. For the present conditions, it is observed that the electron peak density increases from $8.4 \times 10^{15} m^{-3}$ at *N*=1 to $9.4 \times 10^{15} m^{-3}$ at *N*=12 i.e., a ~12% increase is noticed. While, for the same conditions the electron bulk temperature decreases from 1.85 eV at *N*=1 to 1.6 eV at *N*=12.

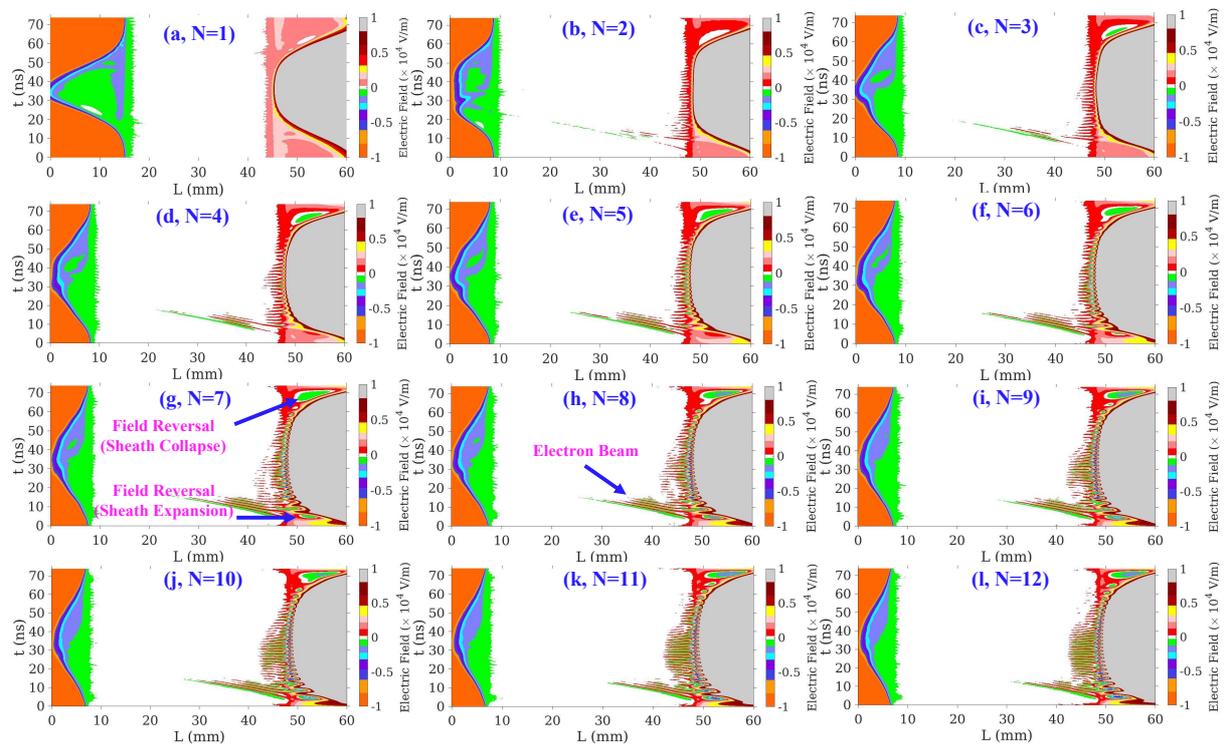

Figure 2. Spatiotemporal electric field distribution in the discharge for different values of *N* varying from 1 to 12.

Figure 3 shows the Fast Fourier Transform (*FFT*) of instantaneous sheath edge position at grounded electrode. The sheath edge position is calculated by observing the maximum position of the electron sheath edge from the electrode where the quasi-neutrality breaks down. In figure 3, the normalized (by fundamental frequency amplitude) values of *FFT* amplitude are plotted for 4 different values of *N* (1 (sine), 4, 8 and 12). As shown in figure 3, for *N*=1 (sine waveform), the higher harmonic on the oscillating sheath is almost negligible accept 2$^{nd}$ harmonic, which is ~20% of the fundamental frequency. As the value of *N* increases, both the number of higher harmonics and its overall contribution with respect to the fundamental increases. For higher values of *N*, it is observed that the 2$^{nd}$ harmonic component increases to ~50%. When comparing the different values of *N*, the frequency of the higher harmonic increases. For *N*=4, it is found that higher order frequencies up to 4$^{th}$ harmonic are present on the sheath oscillations. As the value of $N$ increases to 8, we clearly see the higher harmonics up to $12^{th}$ harmonic in significant percentage. Similar results are observed for $N$=12, with greater contribution of the harmonics shifting on the higher side. Thus, from simulation results it is confirmed that through non-linear interactions increasing number of harmonics in the sawtooth-like waveform appears on the instantaneous sheath edge position. These higher harmonics impart significant effect on the electron/ion heating, which is discussed later in the manuscript.

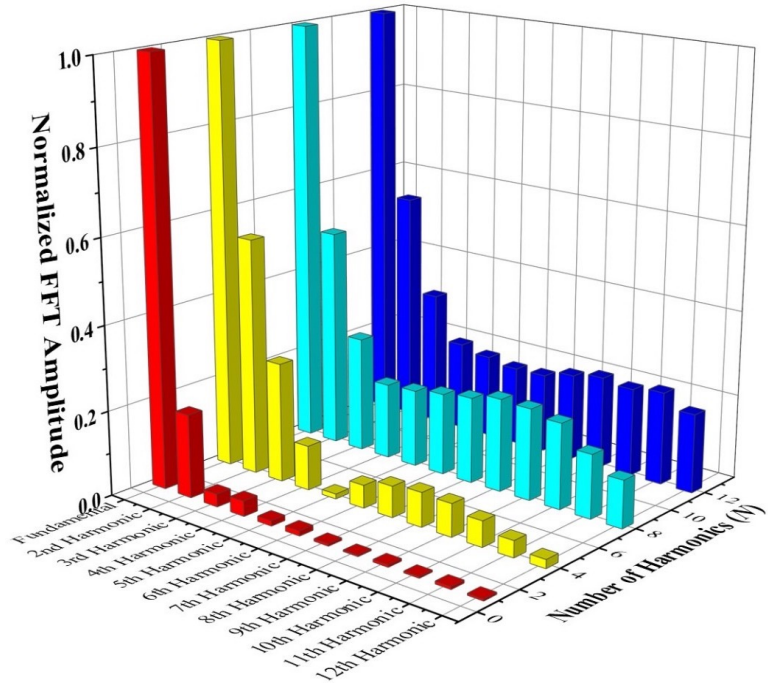

Figure 3. Normalized *FFT* amplitude of the sheath oscillations plotted up to $12^{th}$ harmonic for 4 different values of $N$ i.e. 1 (sine), 4, 8 and 12.

The energy transfer to the electrons after interaction with the varying sheath position and generation/penetration of electron beam is associated with the formation of electrical field reversal i.e., electric field region directing towards bulk plasma. Figure 2 (g, $N$=7) illustrates the field reversal region during the expanding and collapsing phases of an RF sheath for demonstration purposes. Figure 4 shows the surface plot of electric field in the discharge, showcasing four selected values of $N$ (1(sine), 4, 8 and 12). The graphs in the left column (figure 4(a), (c), (e) and (g)) depict the entire discharge system, whereas right hand side column (figure 4(b), (d), (f) and (h)) shows the zoomed region near grounded electrode from center of the discharge. As displayed in figure 4 (a) and 4 (b), for $N$=1 no electric field reversal is observed as the sheaths are smooth with no higher harmonics and therefore no penetration of the electric field in the bulk plasma (also see figure 2(a)). With an increase in the value of $N$ i.e., for $N$=4 (refer to figure 4(c) and (d)), the sheath edge starts to exhibit local time modulation. Nevertheless, the strength of field reversal region is very weak and is not distinctly observed in this case (also see figure 2(d, $N$=4)). In figure 4(e) and 4(f), the surface plot of electric field for $N$=8 is depicted. The presence of the field reversal region during the sheath expansion and other localized field reversal regions is evident (also seen in figure 2 (h, *N*=8)). Additionally, the sheath edge modifications are also noticeable. Significant electric field reversals were observed at *N*=12, reaching magnitudes on the order of $10^4$ V/m. In the case of *N*=12, multiple electric field reversal corresponds to the high frequency oscillation on the instantaneous sheath edge position as displayed in figure 2 (l). The physical phenomenon for the generation of electric field reversal is attributed to the compression of electron fluid during the expanding phase of the sheath edge. During the phase of the sheath expansion, initially, the high energy electrons gain energy and bounce back towards bulk plasma. These electrons after reflection interact with the other slow electrons that are approaching towards the sheath edge and thus electron fluid compresses responsible for the electric field reversal. Multiple electric field reversal is due to the temporal asymmetry of the sawtooth waveform i.e., fast rise and slow fall and electron response time [56, 57, 64]. Figure 5 shows the field reversal strength as a function of *N* during initial expanding phase (i.e., between 0 to 10 ns), its maximum value (between 10 ns to half RF period) and collapsing phase (~ 60 ns to 73 ns). For all cases, the field reversal strength observes to increase with an increase in *N*. For the initial expanding phase of the sheath edge, the field reversal strength increases from ~ (-80 *V/m*) at *N* = 4 to ~ (-4200 *V/m*) at *N* = 12, whereas the maximum field reversal value increases from ~ (-700 *V/m*) at *N* = 4 to ~ (-9400 *V/m*) at *N* = 12. During the collapsing phase, the field reversal strength changes from ~ (-200 *V/m*) at *N* = 2 to ~ (-2000 *V/m*) at *N* = 12. The strength of field reversal has the potential that could lead to significant electron and ion heating, a topic elaborated upon in the subsequent section.

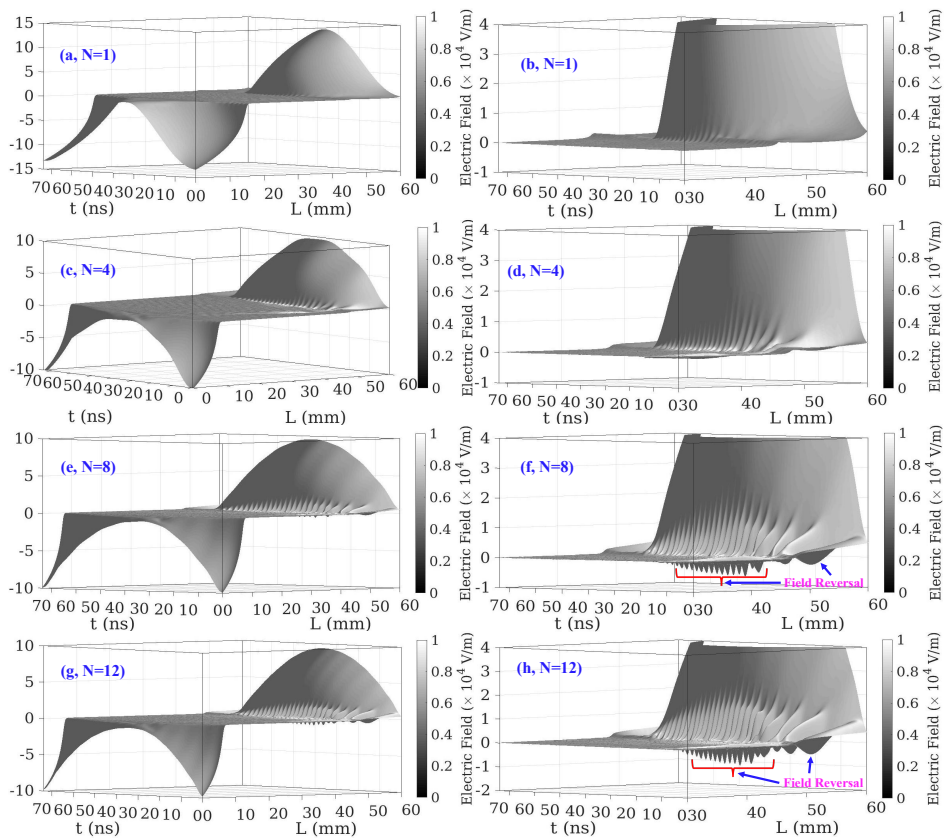

Figure 4. Surface plots of electric field for 4 different values of *N* (1(sine), 4, 8 and 12). Left column shows complete discharge system and right column shows electric field from centre of the centre to grounded electrode. The electric field is clipped at $1\times10^4$ V/m.

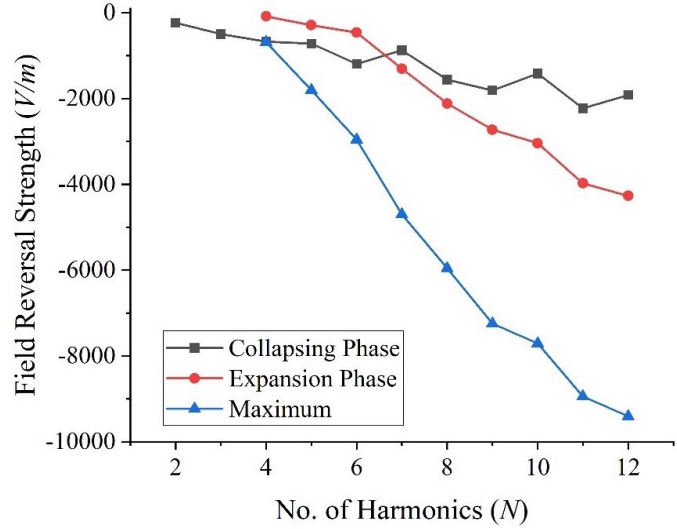

Figure 5. Field reversal strength during expanding phase, collapsing phase and its maximum value as a function of *N*.

Figure 6 shows the charge separation ($n_e - n_i$) in the discharge for different values of *N* i.e. 1(sine), 4, 8 and 12. The normalized value of the charge separation is plotted in figure 6. Once again, the diagrams presented in the first column pertain to the entire discharge system, whereas the figures in the second column are plotted from centre of the discharge to the grounded electrode. The sine waveform (*N*=1) shown in figure 6 (a) and 6 (b) illustrates no charge separation near to the sheath edge i.e., nearly quasi-neutral plasma near to the sheath edge as there is no electric field reversal is present in this case. With the increase in the value of *N*, the charge separation demonstrates the trapping of electrons at locations where field reversal regions are evident in figure 4. In the instance of *N*=4, as shown in figure 6 (c) and (d), electron trapping is minimal due to the relatively weak field reversal regions (as indicated in figure 4 (c) and (d)). However, for *N*=8, as depicted in figure 6 (e) and (f), the presence of trapped electrons is conspicuous owing to the notable field reversal regions observed in this case (as seen in figure 4 (e) and (f)). Ultimately, the charge separation i.e., the trapped electrons attain their peak magnitude at *N*=12 as illustrated in figure 6 (g) and (h), aligning with the most extensive field reversal regions formed in figure 4 (g) and (h).

Multiple charge separation phenomenon occurs near to the phase of sheath expansion due to high frequency oscillation on the instantaneous sheath edge position as displayed in figure 2. The observed charge separation is the trapping and untrapping of the electrons in the electric field reversal region near to the sheath edge. This is the phase of the sheath edge where the electron beams are generated and penetrates in the bulk plasma. This effect is dominant at higher values of *N* where the maximum penetration is observed. The trapping and untrapping of electrons result in the formation of displacement current at the field reversal locations. Figure 7 shows the spatio-temporal and surface profile of displacement current ($I_{DC}$) in the discharge. Figure 7(a) and 7(b) (zoomed part) shows, for *N*=1 (sine), the presence of significant $I_{DC}$, which is mostly confined in the sheath region and there is negligible $I_{DC}$ is present in the bulk plasma or near to the sheath edge. It is also evident in the surface plot of $I_{DC}$ shown in figure 7 (c) where the $I_{DC}$ is not visible except inside sheath region. As the value of *N* increases to 12, significant $I_{DC}$ is observed (in figure 7(d) and 7(e)) at the locations where the field reversal regions formed, leading to trapping and untrapping of electrons. Surface plot of $I_{DC}$ shown in figure 7(f), also shows the formation of multiple $I_{DC}$ at the same locations where the field reversals appear. Thus, these results confirms that the electron trapping/untrapping causes generation of time varying electric field at the multiple field reversals locations.

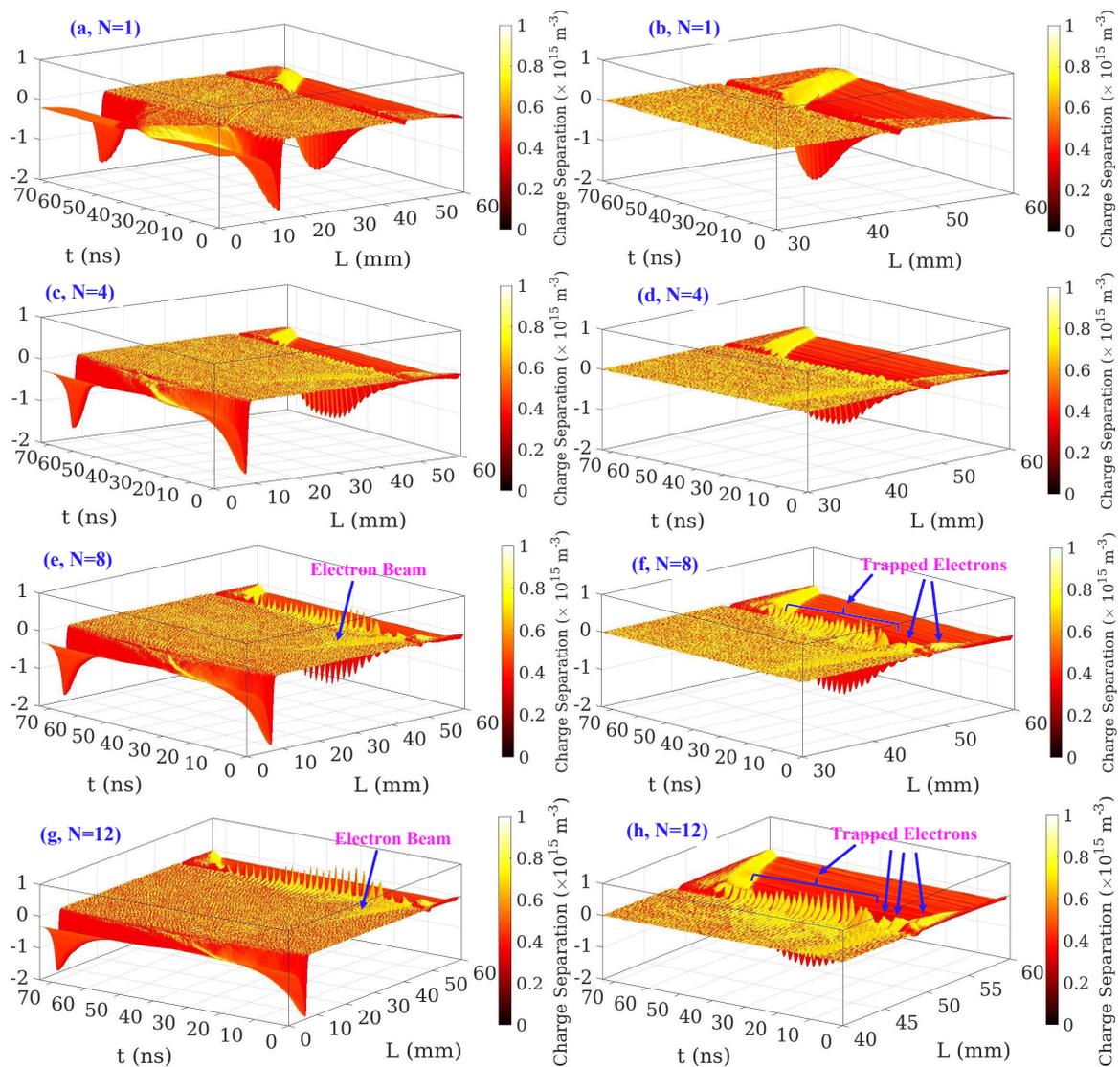

Figure 6. Surface plots of charge separation for 4 different values of *N* (1(sine), 4, 8 and 12).

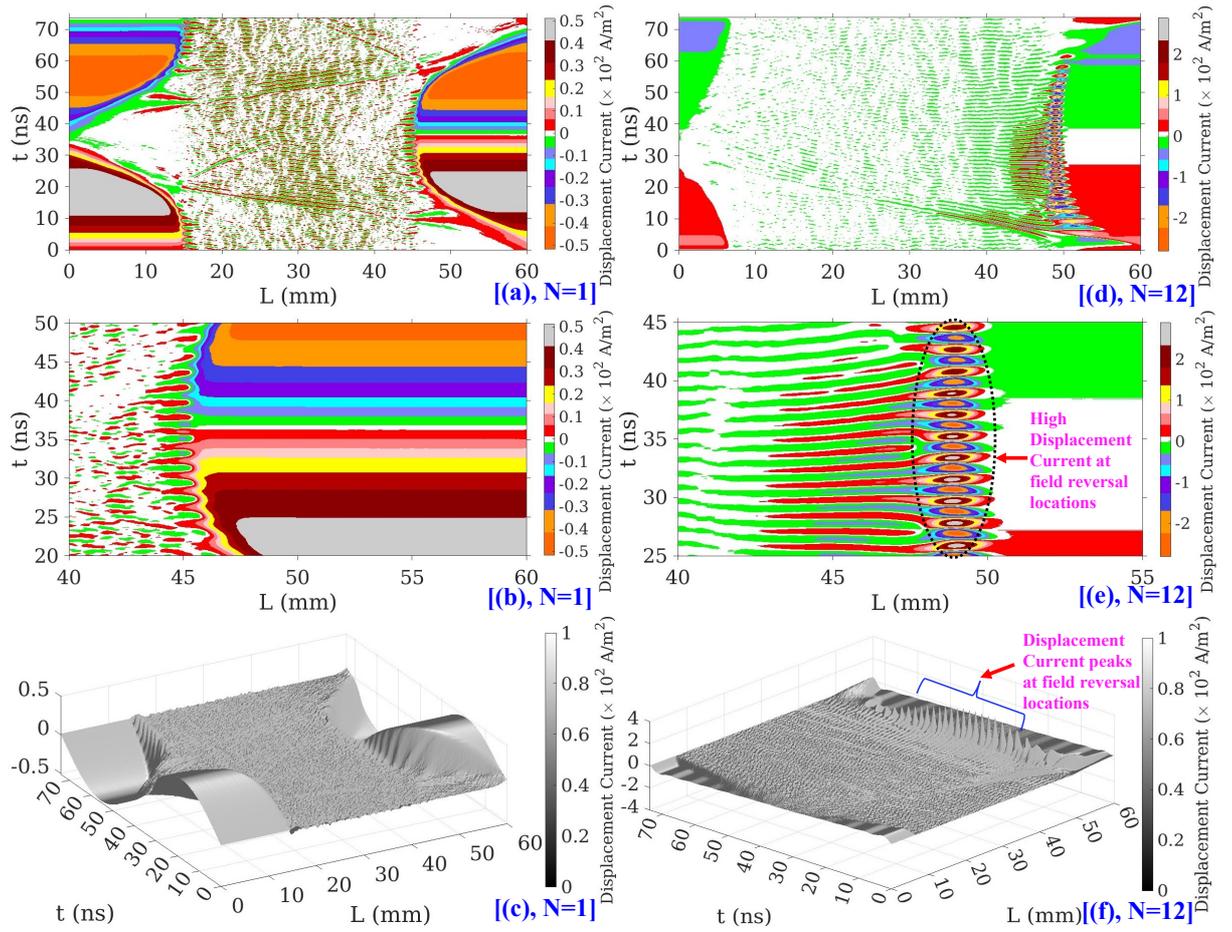

Figure 7. Spatio-temporal profile and surface plot of displacement current in the discharge for $N = 1$(sine) and 12.

Figure 8 shows the electron heating in the discharge for 4 different values of $N$ (1(sine), 4, 8 and 12). As shown in figure 8, the electrons are mostly heated during the expanding phase of the sheath edge. This is the phase where the electric field is growing and sheath velocity increases, which results in the maximum energy transfer to the electrons. It is worth noticing that for $N = 1$ the electron heating is nearly symmetric, and the heating region is broad near to the sheath edge. During this scenario, the most intense heating occurs as the sheath edge expands, while electrons experience energy loss as the sheath collapses. With an increase in the value of $N$, electron heating becomes notably asymmetric, signifying that the maximum heating of electrons is observed near to the grounded sheath edge in comparison to the sheath edge at the powered electrode. This is due to the fact that the applied waveform is temporally asymmetric, which creates highly asymmetric sheath response near to powered and grounded electrodes. Furthermore, the grounded sheath width is larger in comparison to the powered sheath width. Due to an increase in the sheath width/ sheath velocity at the grounded sheath the peak electron heating rate increases from ~2×10$^5$ W/m$^3$ at $N = 4$ to ~14×10$^5$ W/m$^3$ at $N = 12$. In addition, for higher values of $N$ multiple electron heating on the instantaneous sheath edge position is observed, which is due to the fast rise and slow decay of the applied current waveform as shown in figure 1.

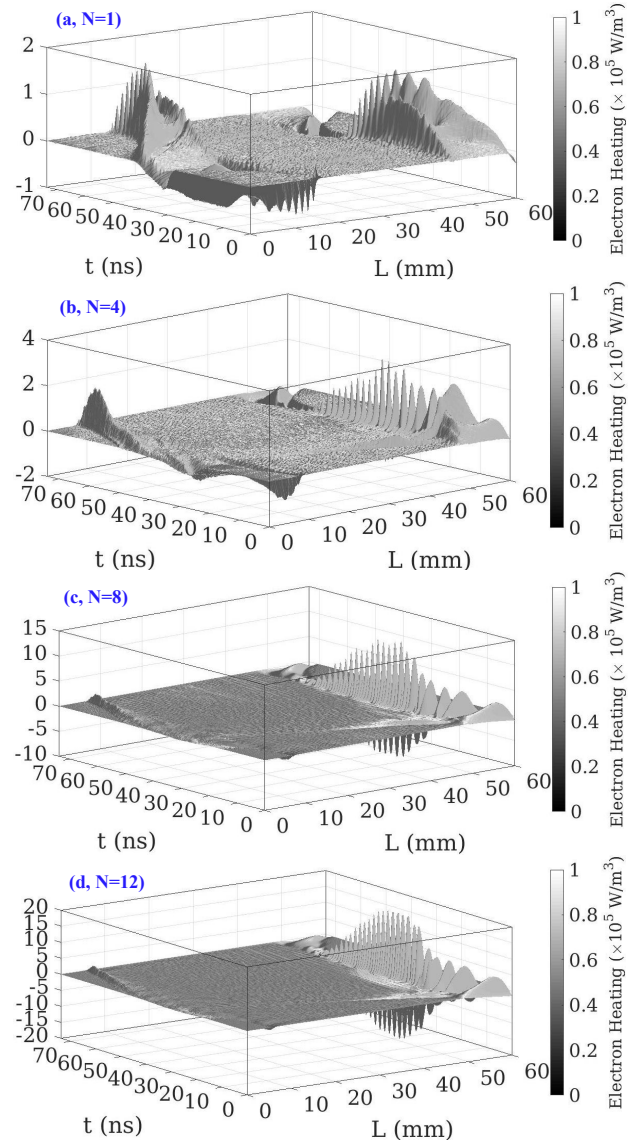

Figure 9. Surface plots of electron heating for 4 different values of $N$ (1(sine), 4, 8 and 12).

Figure 9 shows the surface plots of ion heating in the discharge for different values of $N$ (1(sine), 4, 8 and 12). The left column displays figures for the entire discharge system, while the right column showcases plots from the discharge center to the grounded electrode. From figure 9, it is evident that the ion heating in the bulk plasma is close to zero as there is very weak electric field present there. For sine ($N = 1$), figure 9 (a) and 9 (b), the ions are mostly heated inside the sheath at both powered and grounded electrodes because of the presence of higher electric field. In this case, the ion heating is symmetric and reaching up to a maximum of 15×10$^4$ W/m$^3$ at the electrodes surface. For sawtooth waveforms ($N = 4, 8$ and 12), the ion heating at the electrode surface is reduced to 10×10$^4$ W/m$^3$, however not changing with increasing value of *N*. Moreover, no asymmetry in the ion heating at powered and grounded is observed. In contrary, the ion heating near to the grounded electrode (figure 9 - right column) shows dramatic behaviour. As the value of *N* increases, ion cooling is observed near to the sheath region where the field reversal is present, and electrons are trapped. The ion cooling is highest at $N = 12$, which is nearly $-0.9 \times 10^4$ W/m$^3$. The observed ion cooling is due to the formation of local field reversal regions in the vicinity of the sheath region. It's important to observe that the direction of field reversal region points toward the bulk plasma, leading to its function of accelerating electrons and decelerating ions. At $N=12$, the ion kinetic energy calculations at the sheath edge gives ~7.5 eV based on the ion velocity of ~$6 \times 10^3$ m/s obtained by the simulation. Whereas corresponding average field reversal energy is ~100 eV (calculated by an average of 5000 V/m field reversal strength (figure 5) and 0.02 m average distance travelled by the ion from centre to the sheath edge). Thus, the field reversal strength is strong enough for ion cooling. Again, the alternate ion heating and cooling region corresponds to the high frequency oscillations that is generating the alternate field reversal conditions.

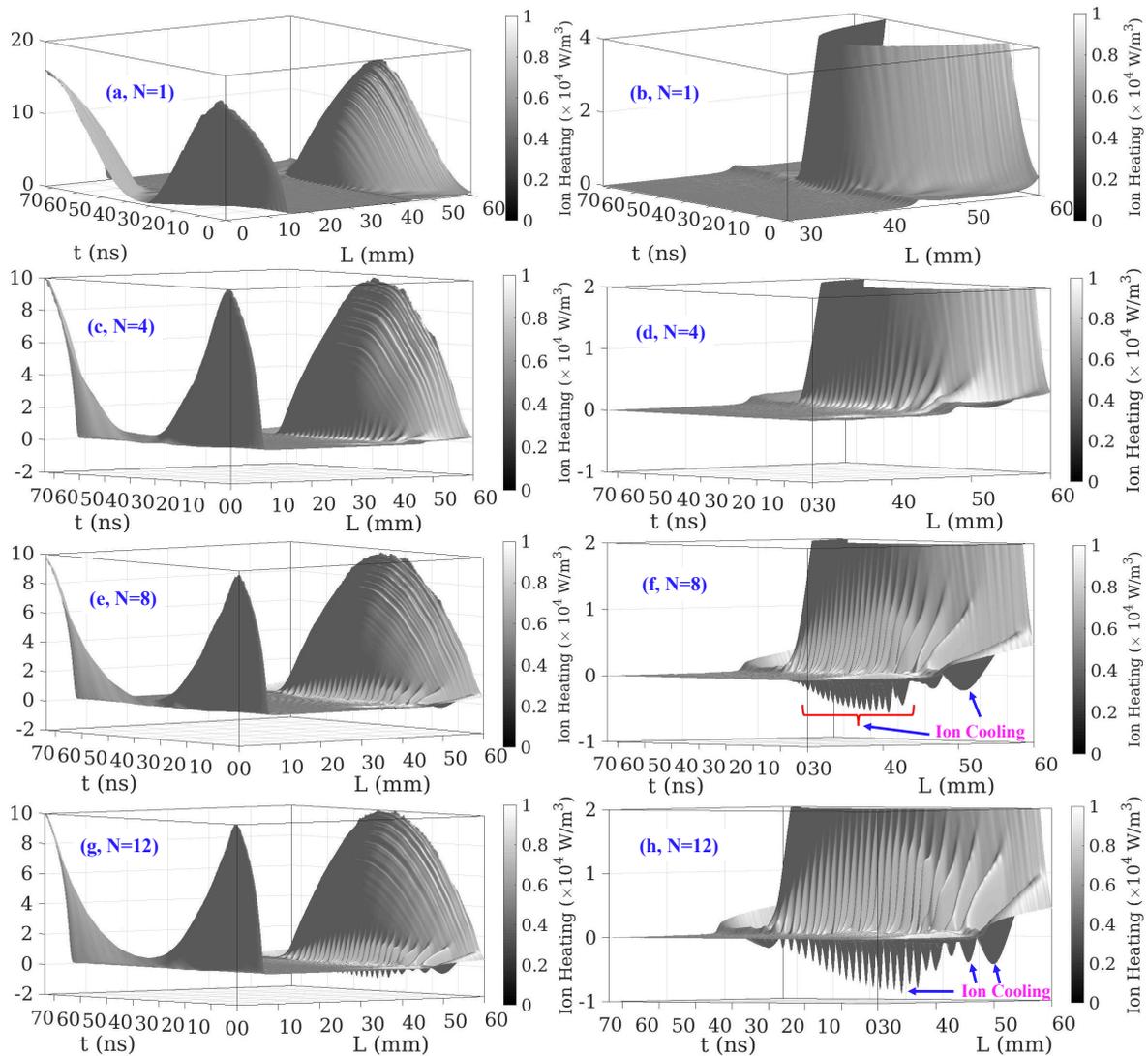

Figure 9. Surface plots of ion heating for 4 different values of $N$ (1(sine), 4, 8 and 12). The ion cooling is clearly visible for higher values of $N$.

## 4. Summary and Conclusions

In summary, using *PIC* simulations we investigated the electric field/sheath nonlinearity, field reversal, charge separation, electron and ion heating in low pressure *CCP* discharges operated by sawtooth-like waveform. A systematic study by varying the number of harmonics ($N$) contained in the waveform is performed in argon plasma. As the value of $N$ increases, the simulation results show a highly nonlinear sheath electric field at the grounded electrode that also penetrates in the bulk plasma. A significant percentage of the higher harmonics up to $12^{th}$ harmonic is observed on the oscillating sheath edge position that are efficient in terms of power deposition and thus plasma density increases. These harmonics are associated with the strong local field reversal and charge separation that is present near to both expanding and collapsing phase of the sheath edge. The field reversal strength is increasing with number of harmonics and found to be greater during the expanding phase, reaching up to ~(-10000 V/m) at $N$=12, which is almost 10 times higher than the field reversal strength during the collapsing phase. This field reversal causes significant electron heating ($15 \times 10^5$ W/m$^3$) and ion cooling ($-1 \times 10^4$ W/m$^3$) at $N$=12 near to the sheath edge. Multiple field reversal regions are observed as the value of $N$ increases analogue to multiple frequency *CCP* discharges. In conclusion, the higher harmonics generation, field reversal, electron trapping/untrapping and displacement current outside sheath region near one electrode is responsible for the plasma asymmetry in the discharges excited by temporally asymmetric sawtooth-like waveform that is beneficial for the plasma processing applications.

**Acknowledgement**

This work is supported by the Science and Engineering Research Board (SERB) Core Research Grant No. CRG/2021/003536.

A. K. is thankful to the Board of Research in Nuclear Sciences (BRNS Sanctioned No. 57/14/04/2022-BRNS), Science and Engineering Research Board EMEQ program (SERB sanctioned No. EEQ/2022/000144) and National Supercomputing Mission (Ref No: DST/NSM/R&D_HPC_Applications/2021/4).